\documentclass[a4paper,fleqn,usenatbib]{mnras}
\usepackage{natbib,geometry, graphicx, psfrag}
\usepackage{mathtools} 
\usepackage{rotating}
\usepackage{amssymb}
\usepackage{amsmath}
\usepackage{color}
\usepackage[utf8]{inputenc}
\usepackage{wasysym}
\usepackage{epstopdf}

\newcommand{\varA}[1]{{\operatorname{#1}}}
\makeatletter

\title[GIs in a Circumbinary Disc]{A 3D Hydrodynamics Study of Gravitational Instabilities in a Young Circumbinary Disc}

\author[K. M. Desai et al.]{
Karna M. Desai,$^{1}$\thanks{E-mail: kardesai@iu.edu}
Thomas Y. Steiman-Cameron,$^{1}$
Scott Michael,$^{1}$
Kai Cai,$^{2}$
\newauthor Richard H. Durisen$^{1}$ \\
$^{1}$Department of Astronomy, Indiana University Bloomington, 727 East 3rd Street, Swain West 318, Bloomington, IN 47405 \\
$^{2}$Pine Manor College, 400 Heath Street, Chestnut Hill, MA 02467}

\date{Accepted XXX. Received YYY; in original form ZZZ}

\pubyear{2018}

\begin{document}
\label{firstpage}
\pagerange{\pageref{firstpage}--\pageref{lastpage}}
\maketitle

\begin{abstract}
We present a 3D hydrodynamics study of gravitational instabilities (GIs) in a 0.14 M$_{\odot}$ circumbinary protoplanetary disc orbiting a 1 M$_{\odot}$ star and a 0.02 M$_{\odot}$ brown dwarf companion. We examine the thermodynamical state of the disc and determine the strengths of $\varA{GI-induced}$ density waves, nonaxisymmetric density structures, mass inflow and outflow, and gravitational torques. Results are compared with a parallel simulation of a protoplanetary disc without the brown dwarf binary companion. Simulations are performed using CHYMERA, a radiative 3D hydrodynamics code. The onset of GIs in the circumbinary disc is much more violent due to the stimulation of a strong $\varA{one-armed}$ density wave by the brown dwarf. Despite this early difference, detailed analyses show that both discs relax to a very similar $\varA{quasi-steady}$ phase by 2,500 years after the beginning of the simulations. Similarities include the thermodynamics of the $\varA{quasi-steady}$ phase, the final surface density distribution, radial mass influx, and nonaxisymmetric power and torques for spiral arm multiplicities of two or more. Effects of binarity in the disc are evident in gravitational torque profiles, temperature profiles in the inner discs, and radial mass transport. After 3,800 years, the semimajor axis of the binary decreases by about one percentage and the eccentricity roughly doubles. The mass transport in the outer circumbinary disc associated with the $\varA{one-armed}$ wave may influence planet formation.

\end{abstract}




%
\section{Introduction}  
\label{sec:intro}
%

%
%

A circumbinary disc is a disc around two stars orbiting the common center of mass, and a circumbinary exoplanet is a planet orbiting both stars in a binary pair. Circumbinary exoplanets presumably form in circumbinary discs. Over 3,800 exoplanets have been discovered as of November of 2018\footnote{\label{exo_archive}NASA Exoplanet Archive, http://exoplanetarchive.ipac.caltech.edu}, of which 21 are circumbinary exoplanets.  
See \citet{Haghighipour2010} for a comprehensive review of planets in binary stars. 

The question we address in this paper is how the presence of a central binary may affect the evolution of a circumbinary protoplanetary disc in the case where the disc is massive and subject to gravitational instabilities (GIs). We use a 3D radiative hydrodynamics code with realistic dust opacities to produce two simulations for the same gas disc. In one simulation, the disc orbits a single $\varA{solar-type}$ star. In the other, the $\varA{solar-type}$ star is given a relatively close brown dwarf companion. We then compare the onset and subsequent development of GIs.

GIs have long been recognized as a mechanism for producing angular momentum and mass transport in astrophysical discs \citep{Kuiper51,toomre1964,goldreich1965,Pringle81}. GIs probably occur during the early phases of protostellar disc evolution when the disc is likely to be massive \citep{Adams1993}. For comprehensive reviews of GIs in protoplanetary discs, please refer to \citet{Durisen2007} and \citet{Kratter16}. 

The Toomre $Q$ parameter
\begin{equation}
Q = \frac {\kappa {\rm c_s}} {\pi G \Sigma},
\end{equation} 
where ${\rm c_s}$ is the sound speed, $\kappa$ is the epicyclic frequency, and $\Sigma$ is the disc surface mass density, is often used to measure a disc's susceptibility to GIs \citep{toomre1964}. For $Q$ $\lesssim$ 1.7, small density perturbations can grow exponentially due to GIs in the linear regime on a timescale comparable to the local dynamical time \citep{Laughlin98,Nelson98,Pickett98}. The precise $Q$ stability limit for nonaxisymmetric modes depends in detail on the structure of the disc \citep[][and references therein]{Durisen2007}. In a $\varA{GI-active}$ disc, the density perturbations develop into trailing spiral waves over a wide range of radii \citep{Papaloizou1991,Pickett98,Pickett00,Pickett03,Laughlin98,Nelson98,Nelson00}. The spiral waves serve to heat the disc, thus increasing $Q$ and decreasing the disc's susceptibility to GIs. At the same time, radiative cooling acts in the opposite direction, driving the disc towards instability. Sufficiently rapid cooling of the disc may lead to fragmentation \citep{Boss2001,Gammie2001,Boss2002,JohnsonGammie2003,Rice2005,Boss06bin,Baehr2017}. 
When cooling is not too rapid, a balance can be reached between heating and cooling, resulting in a relatively constant, but unstable, value of $Q$. When this occurs, the disc exists in a $\varA{long-lived}$, $\varA{quasi-steady}$, marginally unstable state, and the GI activity leads to sustained mass and angular momentum transport \citep{goldreich1965,paczynski1978,LP87,Gammie2001,LodatoRice2004,mejia2005,Boley2006-2,Durisen2007,Michael12,SteimanCameron2013}.

This paper investigates the extent to which the presence of an inner brown dwarf companion affects GIs, especially once the disc has settled into a $\varA{quasi-steady}$ marginally stable state. Preliminary results of these simulations were reported in \citet{Cai2013}. Numerical methods employed in this study and the simulations performed are described in Section \ref{sec:methods}. Major results, discussion, and conclusion are presented in Section \ref{sec:bin_res}, Section \ref{sec:discussion}, and Section \ref{sec:conclusion}, respectively.

%
\section{Methods} 
\label{sec:methods}
%
\subsection{3D Hydrodynamics Code} 
\label{sec:hydrocode}
%
We use the $\varA{grid-based}$ 3D hydrodynamics code CHYMERA (Computational HYdrodynamics with MultiplE Radiation Algorithms) to perform the simulations. CHYMERA is an Eulerian, 2$^{nd}$ order accurate code both in space and time that solves the Poisson equation and the equations of hydrodynamics on a uniform cylindrical $(r,\phi,z)$ grid, including an energy equation with PdV work, net heating or cooling due to radiative flux divergence, and heating by artificial bulk viscosity. The code assumes mirror symmetry about the equatorial plane or the disc midplane. CHYMERA evolved over the years through contributions from many authors \citep{tohline1980,Yang92,Pickett95,Pickett03,M04,mejia2005,CaiT06,BoleyT07,MichaelT11}.

In this work, the radiative cooling scheme of \citet{Boley2007a} is adopted. This uses $\varA{flux-limited}$ diffusion for optically thick regions in the $r$ and $\phi-$directions and a $\varA{single-ray}$ $\varA{discrete-ordinate}$ radiative transfer solver in the $z-$direction that treats both optically thick and thin regions. The opacity tables are obtained from \citet{DAlessio2001}. To determine the opacities, minimum and maximum grain sizes of 0.005 $\mu$m and 1 $\mu$m are used, respectively. The dust size distribution follows a 
\begin{equation}
n(a) = n_o a^{-s}
\end{equation} 
$\varA{power-law}$ relation, where $n_o$ is the normalization constant, $a$ is the grain radius, and $s$ has the value 3.5 for the current work. The ideal gas equation of state includes the internal degrees of freedom for $H_2$ \citep{Boley2007b}, and the ortho/para ratio of hydrogen is 3:1.

%
\subsection{Initial Models and Simulations} 
\label{sec:sims}
%
The simulations follow the evolution of a 0.14 M$_{\odot}$ disc surrounding a central 1 M$_{\odot}$ star. The initial surface density profile follows the 
\begin{equation}
\Sigma(r) = \Sigma_o r^{-p}
\end{equation} 
$\varA{power-law}$ relation, with $p$ = $0.5$. We use the method described in \citet{mejia2005} to produce the initial physical and thermodynamical conditions of the disc. The simulation of the protoplanetary disc around a single star will be called the ``control'' simulation and was already reported  by \citet{MichaelL11}. The ``circumbinary'' simulation is identical except for the presence of a 0.02 M$_{\odot}$ brown dwarf companion initially orbiting the central star at a radius of 2.5 AU. The simulations reported here are computed using (512,512,64) grid elements in the ($r$,$\phi$,$z$) directions, where the $z-$axis is the rotation axis. Each $\varA{grid-cell}$ is $\approx \frac{1}{6}$ AU in both $r$ and $z$ directions. The inner boundary of the initial disc is at 5 AU, thus creating an inner hole in the disc with no mass. The outer radius of the initial disc is 40 AU. 
Outflow boundary conditions are enforced at the upper vertical grid boundary, the outer radial grid boundary, and an inner radial boundary at 2 AU. An initial $0.01\%$ random $\varA{cell-to-cell}$ perturbation is added to the density distribution to generate nonaxisymmetry. Both the control and the circumbinary cases are simulated for 21 ORPs $\approx$ 3,800 years, where 1 ORP = the initial outer rotation period at 33 AU ($\approx$ 179 years).

While the star remains fixed at the grid center for computational convenience, we account for the star's acceleration through the indirect potential method \citep{AdamsRuden1989,NelsonPapaloizou00}, as discussed in \citet{Michael10}. In this scenario, the reference frame of the star plus grid is accelerated through inclusion of fictitious forces that account for the gravitational interactions between the star and the disc and between the star and the brown dwarf. The orbital motion of the brown dwarf companion is integrated by a Verlet method \citep{Hut95}. Gravitational interactions of the brown dwarf with the disc and star are fully included, and the acceleration of the star by the brown dwarf is also treated by an indirect potential, as in \citet{MichaelL11}. 

The masses of the star and brown dwarf and the orbital distance are adopted to mimic the HD 202206 system \citep{Correia_2005}, where the central 1.1 M$_{\odot}$ star has two companions at 0.83 AU and 2.55 AU with lower mass limits of 17 M$_{\jupiter}$ and 2.4 M$_{\jupiter}$, respectively (where M$_{\jupiter}$ = Jupiter mass). Our choice of brown dwarf mass 0.02 M$_{\odot}$ is approximately 21 M$_{\jupiter}$, which roughly resembles that of the more massive companion of the HD 202206 system.

\begin{figure*}
       \includegraphics[width=0.65\textwidth]{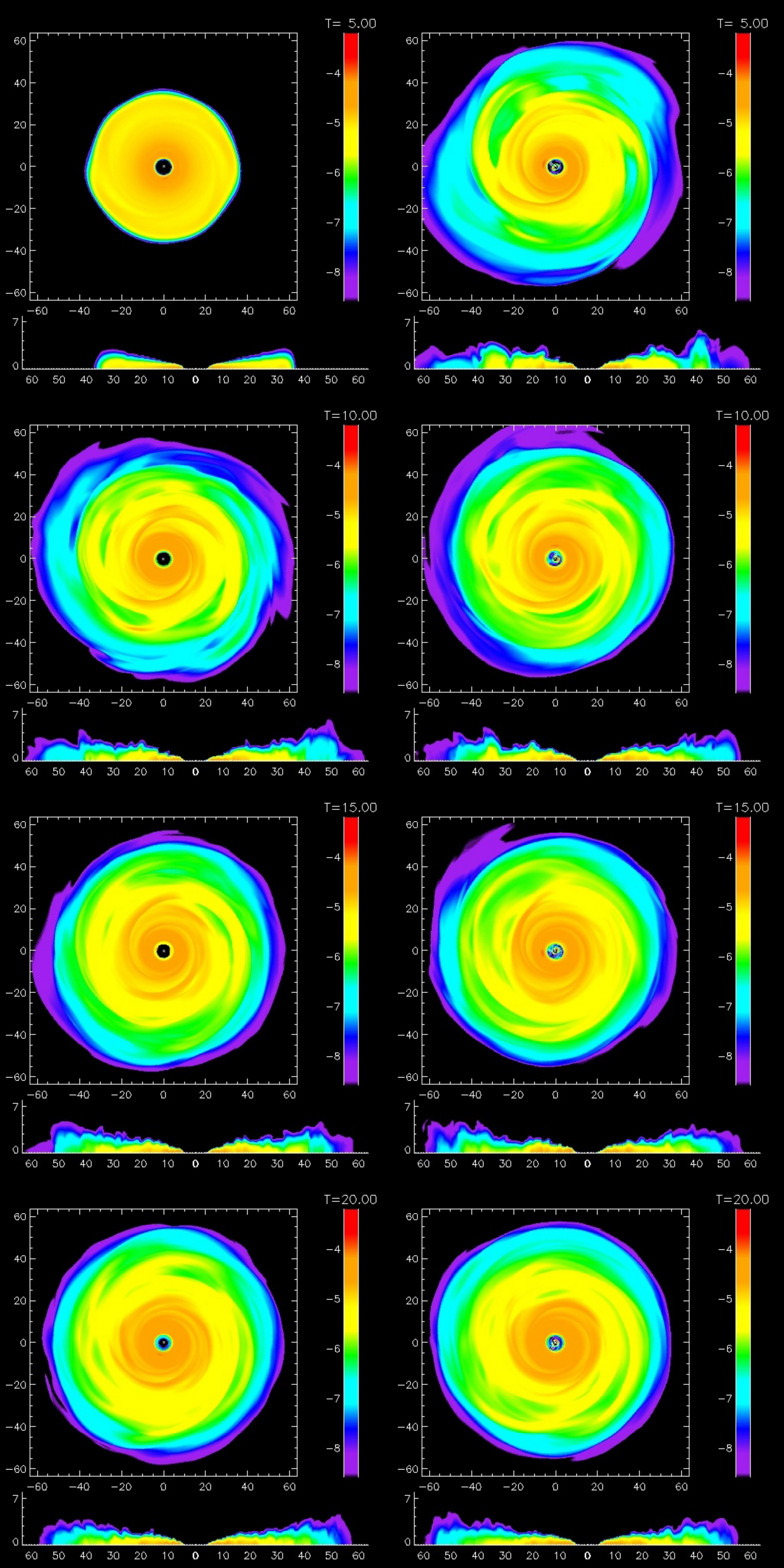}
\caption{\small{Midplane and meridional density contours, in program units, on a logarithmic scale for control (left panels) and circumbinary (right panels) simulations, at 5, 10, 15, and 20 ORPs from top to bottom. The rotational flow is counterclockwise. 1 ORP = Initial Outer Rotation Period at 33 AU ($\sim$179 years). $\varA{Panel-dimensions}$ are in AU.}}
\label{fig:midplane_2017}
\end{figure*}
\begin{figure*}
        \centering
       \includegraphics[width=1\textwidth]{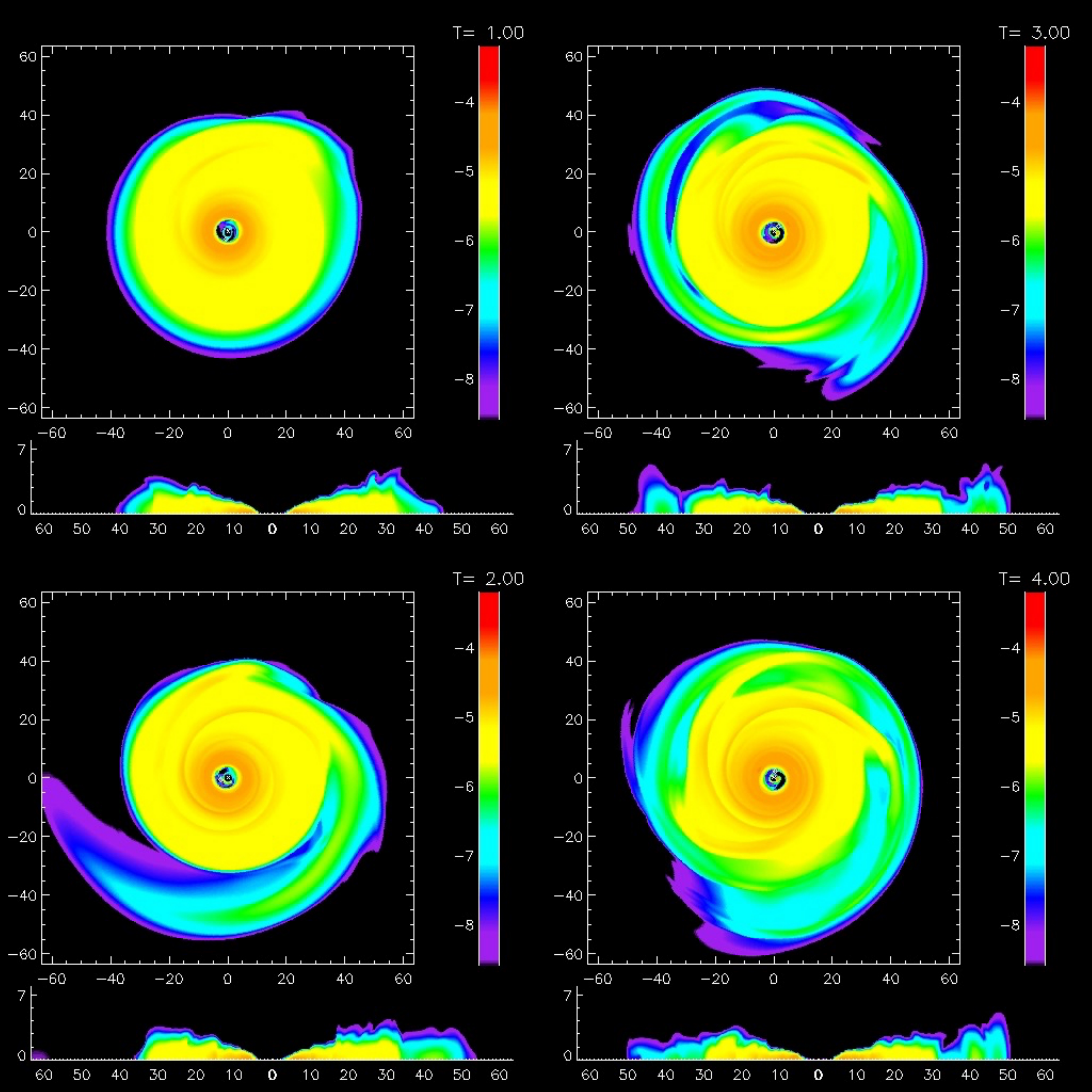}
\caption{Midplane and meridional density contours for the circumbinary simulation at 1, 2, 3, and 4 ORPs from top to bottom and left to right. The figure description is similar to Fig. \ref{fig:midplane_2017}.}
\label{fig:midplane_2017_2}
\end{figure*}
 
\section{Results} 
\label{sec:bin_res}
%
Fig. \ref{fig:midplane_2017} shows the midplane and the meridional density contours for both simulations over a range of times. Similarities and differences between the structure and physical states of the simulations, both for early and late stages, are presented in the subsections below. 

%
\subsection{Early Evolution}  
\label{sec:bin_res_early}
%

The early evolution, although a transient effect, is very different in the two discs. Differences can be observed for approximately the first 12 ORPs (about 2,150 years), but are dramatic over the first few ORPs. When compared with the circumbinary simulation, the control simulation has a relatively slow onset of $\varA{GI-activity}$. In the circumbinary disc, the introduction of the brown dwarf companion changes the central gravitational potential of the system, introducing a relatively strong initial $m = 1$ perturbation. Therefore, the onset of GIs is hastened. The $\varA{GI-activity}$ then rapidly spreads throughout the entire circumbinary disc. The spread begins with the growth of a strong $\varA{one-armed}$ wave. Fig. \ref{fig:midplane_2017_2} illustrates the outward propagation and growth of the $m = 1$ wave in the circumbinary disc at early times and the appearance of other strong $\varA{low-order}$ spiral waves.

\begin{figure*}
        \centering
                \includegraphics[width=0.8\textwidth]{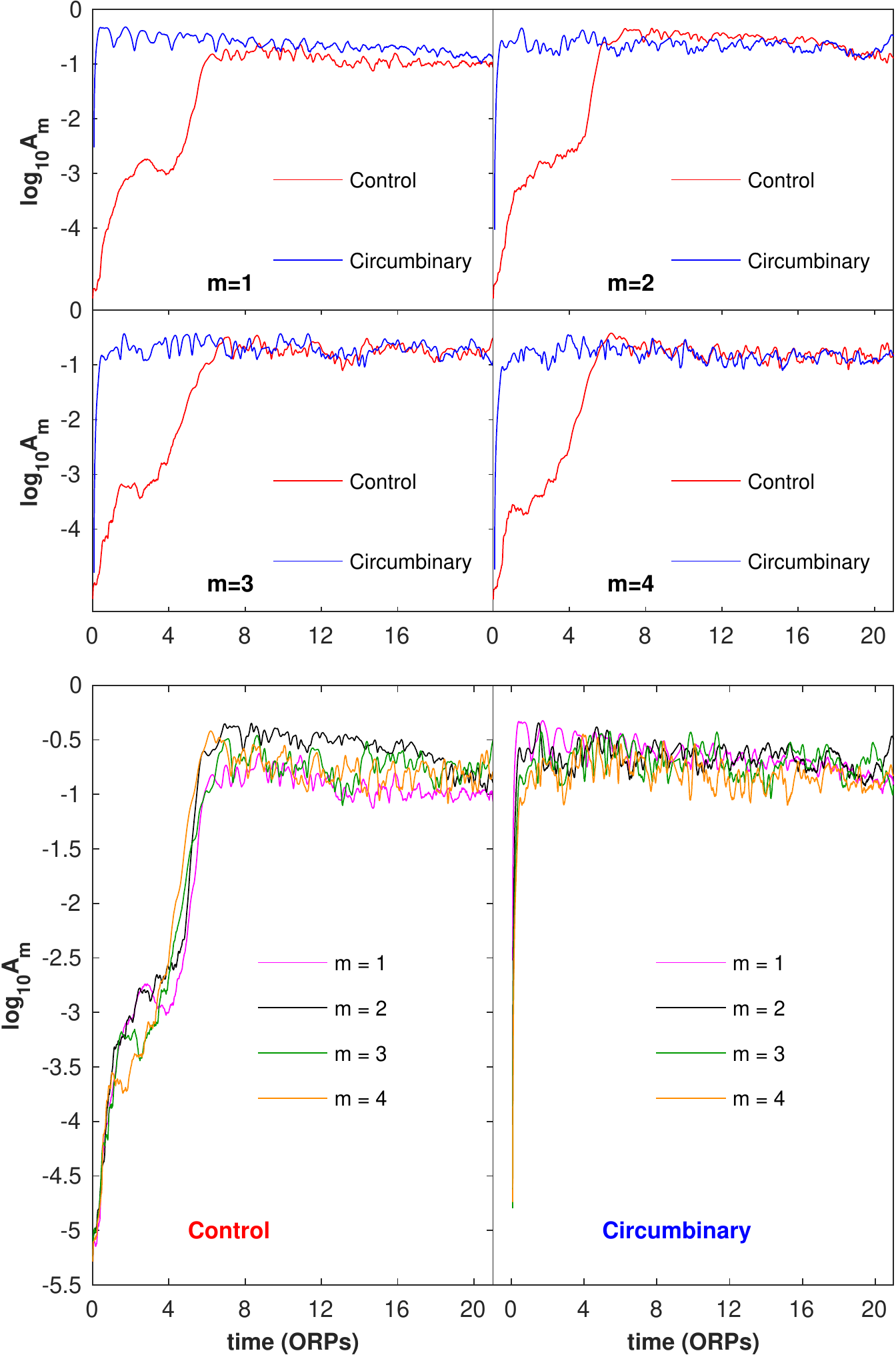}
  \caption{Comparison of Fourier amplitudes  $A_m$, derived from a Fourier decomposition of the azimuthal density distribution, as a function of time for $m$ = 1 to 4, computed over the range 10 to 40 AU, for both discs.}
\label{fig:Ams}
\end{figure*}

By about 5 ORPs, as shown in Figs \ref{fig:midplane_2017} and \ref{fig:midplane_2017_2}, the entire circumbinary disc is $\varA{GI-active}$. Contrarily, the control disc becomes $\varA{GI-active}$ by about seven ORPs, and, as shown in Fig. \ref{fig:Ams} (bottom panel), the burst of GIs begins first in $\varA{high-order}$ disturbances, as can be seen by $m = 4$ structure growing exponentially before any other.

Eventually, strong $\varA{few-armed}$ spiral structures develop through the entire disc in both cases.
This initial growth is referred to as the ``burst'' phase \citep{mejia2005}, a transient that is sensitive to the initial conditions. This phase lasts for a few to several ORPs. After it ends, the disc takes about 1,200 years to settle into $\varA{quasi-steady}$ behavior. The burst is a period of heating and rapid rearrangement of material in the disc. This onset for GIs in simulations of marginally gravitationally unstable discs around a central star has been described in \citet{mejia2005}, \citet{Cai2006}, \citet{Boley2006-2}, and \citet{Michael12}.

This notable difference in behavior of the control and circumbinary simulations at early times can be quantified through a Fourier decomposition of the density structure in $\sin(m\phi)$ and  $\cos(m\phi)$, where $\phi$ is the azimuthal coordinate and $m$ represents the number of arms for cases where coherent spiral waves are produced. We can define a single global parameter $A_m$ that characterizes the relative amount of the disc's mass involved in disturbances with $m-$armed symmetry by

\begin{equation}
A_m =\frac { { ({a_m}^2 + {b_m}^2 })^{ \frac { 1 }{ 2 }  } }{ \pi \int { {\rho}_{0} {\rm rdrdz} }  }, 
\end{equation} 
where $\rho_{ 0 }$ is the axisymmetric component of the density, and 

\begin{equation}
a_m=\int { \rho \cos({m}\phi ) } {\rm rdrdzd}\phi, 
\end{equation} 

\begin{equation}
b_m=\int { \rho \sin({m}\phi ) } {\rm rdrdzd}\phi. 
\end{equation} 

\begin{figure*}
        \centering

                \includegraphics[width=\textwidth]{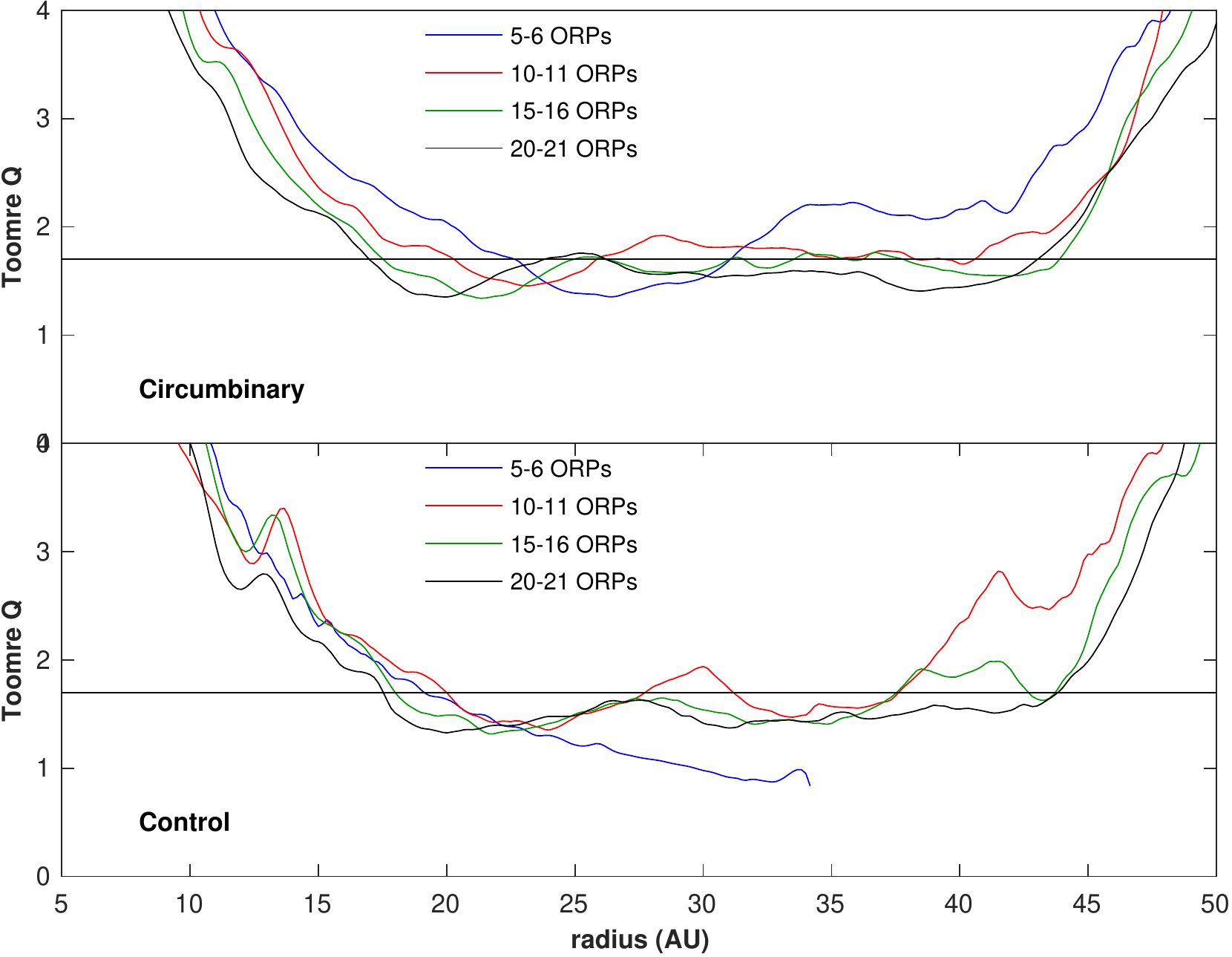}
  \caption{Azimuthally averaged Toomre $Q$ parameter for a range of times. At $5-6$ ORPS, the control disc does not extend beyond 34 AU. Horizontal lines are plotted at  $Q = 1.7$ in both panels.}
\label{fig:Qtoomre}
\end{figure*}
\begin{figure*}
        \centering

                \includegraphics[width=0.65\textwidth]{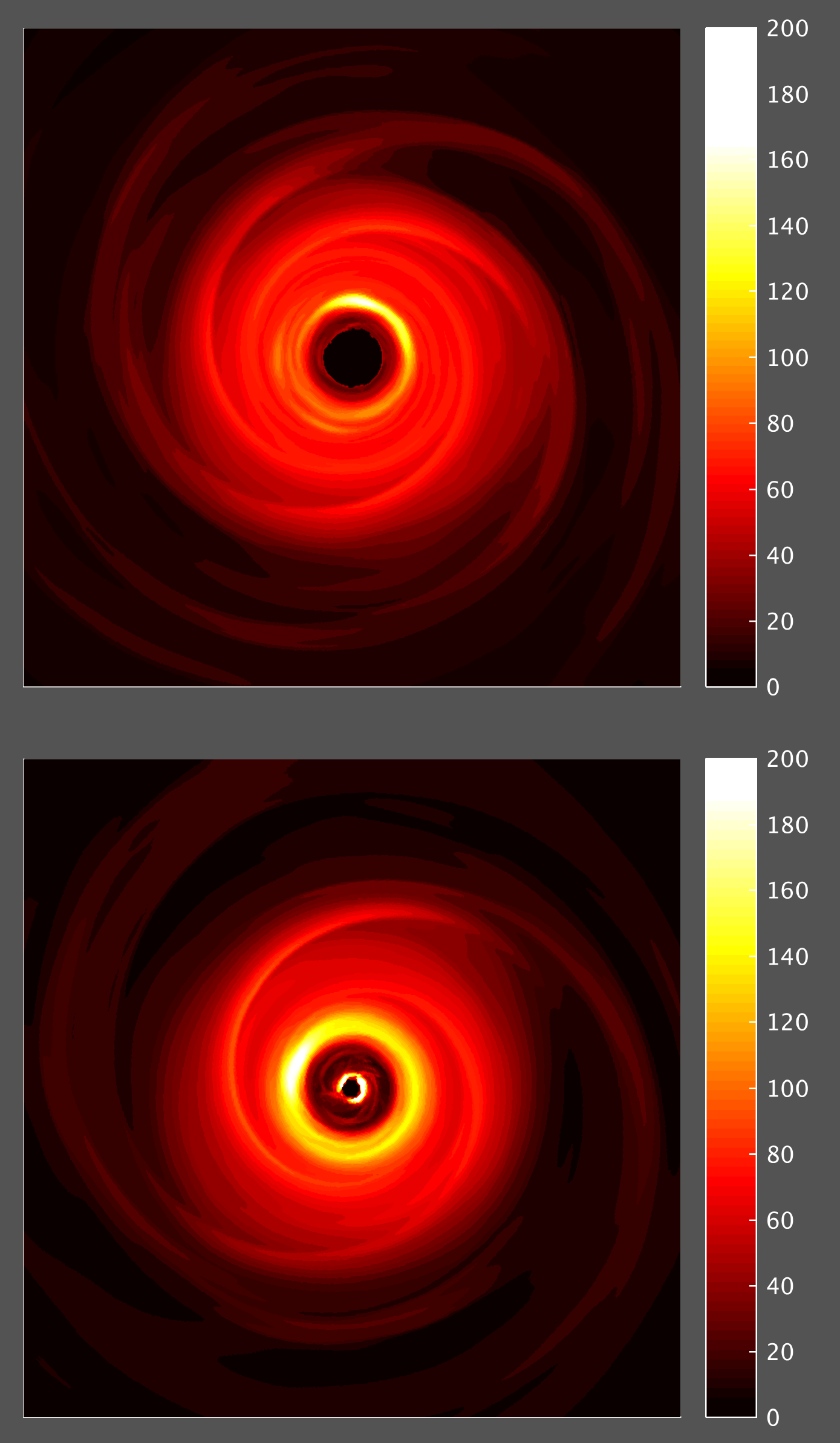}
  \caption{Comparison of temperatures (in Kelvin) in the midplane of the control disc (top panel) and the circumbinary disc (bottom panel) at 21 ORPs. Each panel is 100 AU by 100 AU.}
\label{fig:temperature_comparison}
\end{figure*}

Fig. \ref{fig:Ams} illustrates the Fourier amplitudes, which show significant differences in the early evolution described above. In addition to the abrupt onset of GIs in the circumbinary case, the bottom panel of Fig. \ref{fig:Ams} shows that $A_1$ tends to be greater than or comparable to the other ${A_m}$'s throughout the simulation. For the control case, on the other hand, $m = 2$ tends to be the dominant Fourier component throughout the nonlinear regime, while $m = 1$ is relatively unimportant. At later times, the simulations have more similar distributions of amplitudes, except for $m = 1$. 

Fig. \ref{fig:Qtoomre}, which shows the evolution of the Toomre $Q$ parameter in the two simulations, illustrates another effect of the different ways that GIs set in. By 5$-$6 ORPs, the circumbinary disc has already expanded its radius considerably and been heated by strong spiral waves. The control disc, on the other hand, is still cooling and in fact even shrinking radially at this time, because strong GI waves have not yet erupted.

The differences between the simulations at early times suggest that initiation of $\varA{GI-activity}$ in a circumbinary disc could be relatively violent and be initially dominated by $\varA{one-armed}$ structure. It is difficult to know, however, how this might apply to real systems where the discs may transition more slowly to a marginally unstable state. Despite the early violence of the $\varA{GI-onset}$, there is no hint of fragmentation, and the circumbinary disc ultimately relaxes into a $\varA{quasi-steady}$ state similar to that seen in the control simulation.
%
\subsection{Quasi$-$Steady State}  
\label{sec:quasisteady}
%
In the earlier works of \citet{Pickett03}, \citet{mejia2005}, \citet{Boley2006-2}, \citet{Cai2006}, \citet{MichaelT11}, and \citet{SteimanCameron2013}, after about 10 to 12 ORPs, the discs asymptotically approach a $\varA{quasi-steady}$ state where heating by $\varA{GI-activity}$ is roughly balanced by radiative cooling. This state is characterized by relatively constant, marginally unstable values of $Q$, relatively steady amplitudes of the nonaxisymmetric spiral wave structure, significant torques within the disc dominated by a few $\varA{low-order}$ global spiral modes, and a resulting significant radial mass flow in the disc due to outward angular momentum transport. As shown in Figs \ref{fig:Ams} and \ref{fig:Qtoomre}, there are significant fluctuations about this average state both in time and space. 

These results show that, apart from some differences in dominant spiral waves, especially due to $m = 1$, the circumbinary disc achieves a $\varA{quasi-steady}$ state similar to that of the control disc, even though the onset of GIs in the first 10 ORPs is significantly different. The next few sections lay  this out in quantitative detail.

%
\subsubsection{Thermodynamic State}  
\label{sec:thermodynamics}
%

Fig. \ref{fig:Qtoomre} illustrates that during the time interval of 5 to 6 ORPs, the $Q$ values are lower in the control disc from about $25-35$ AU. At these times, the control disc is still fairly axisymmetric, as shown in Figs \ref{fig:midplane_2017} and \ref{fig:Ams}. The control disc undergoes several ORPs of initial cooling before the spiral waves reach amplitudes above 10 percent. As mentioned in \ref{sec:bin_res_early}, the disc does contract slightly in radius, from 40 to 35 AU as it cools. The circumbinary disc, on the other hand, has already erupted into GI activity and expanded radially due to the passage of strong $\varA{low-order}$ spiral waves.

Differences persist in both discs for different radial regions of the discs when we compare later times of $10-11$ ORPs and $15-16$ ORPs. For example, in the region around $13-15$ AU, the circumbinary disc has a smooth decreasing $Q$ profile, whereas the $Q$ profile in the control disc has a local peak in $Q$ in the same region. A local peak for $10-11$ ORPs in the radial region around 30 AU is seen in both discs. 

Discounting these peaks during $10-11$ ORPs, both discs appear to be approaching the $\varA{quasi-steady}$ phase by 10 ORPs given the fluctuations typical of the $\varA{quasi-steady}$ state. Although the outer control disc does have a significantly higher initial $Q$ profile, it steadily decreases within few ORPs as well. 

Once the discs have settled, $Q$ values fluctuate about a relatively constant $Q$ $\approx$ 1.5 to 1.7 over the radial range $15-45$ AU, which is consistent with the discs being in a $\varA{quasi-steady}$ state of marginal instability. Differences in $Q$ at a given radius between the simulated discs are not significant starting around 14 ORPs, and the $Q$ profiles after 14 ORPs are characteristic of the asymptotic $\varA{quasi-steady}$ state for both simulations. Therefore, we consider both discs to have reached a $\varA{quasi-steady}$ state of balanced heating and cooling for the duration of 14 to 21 ORPs. It is then possible to use averages over these times to compare the asymptotic states.

Fig. \ref{fig:temperature_comparison} compares the temperature profiles in both discs at the end of simulations. Temperatures in the discs are very similar except for the inner regions, where the circumbinary disc is considerably hotter. The presence of the brown dwarf appears to produce additional heating in the inner disc. There is some material near the brown dwarf that gets rather hot, but it represents a vanishingly small amount of mass.

%
\subsubsection{Nonaxisymmetric Structure}  
\label{sec:nonaxisymmetric}
%
\begin{figure*}
        \centering
                \includegraphics[width=\textwidth]{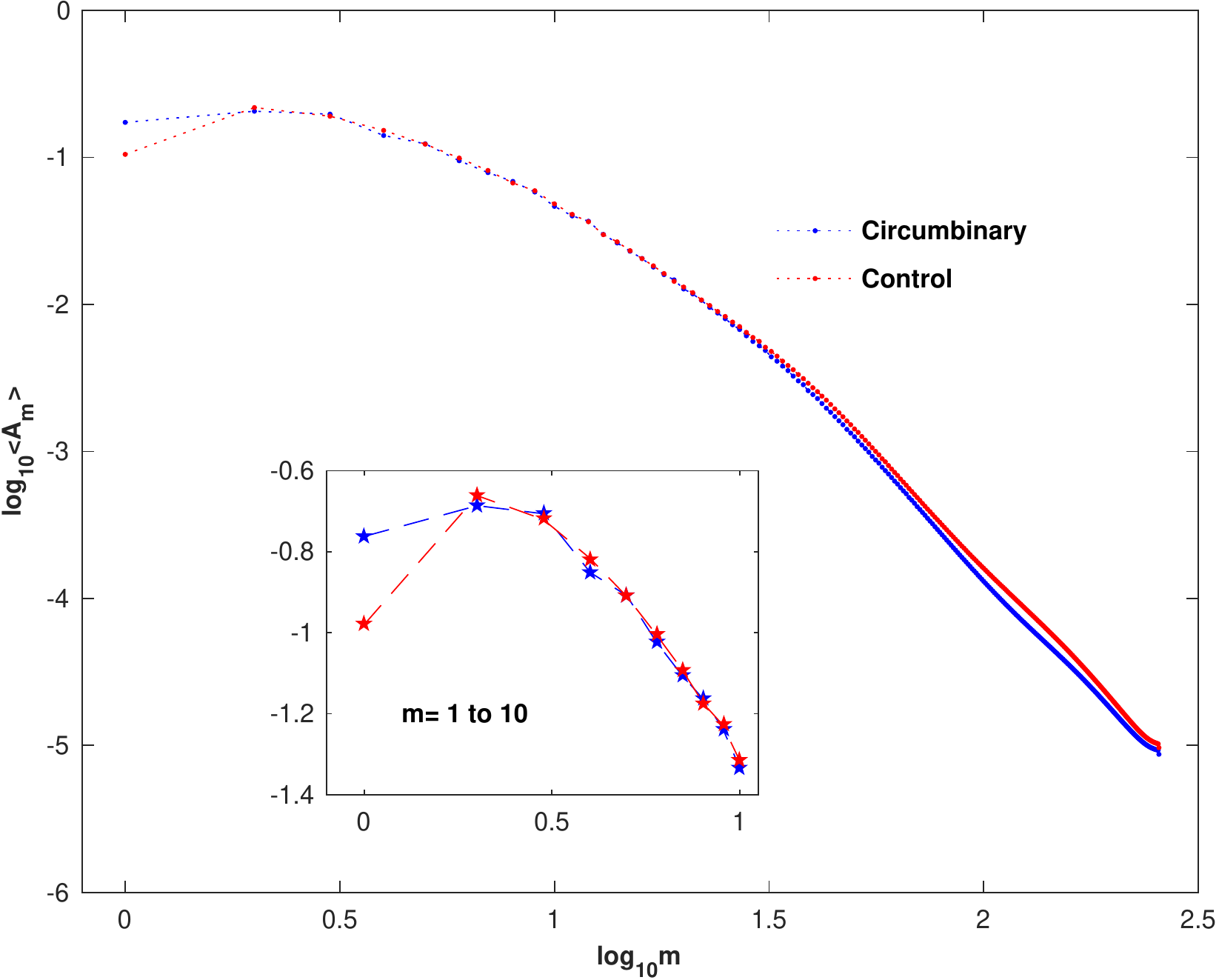}
  \caption{$\varA{Time-averaged}$ Fourier amplitudes $\langle A_m \rangle$ calculated over the radial range 10 to 40 AU and averaged over 14 to 21 ORPs. Inset: Zooming in on the $\langle A_m \rangle$ values for $m=$ 1 to 10. The brackets around $A_m$ indicate that the plotted values have been time-averaged.}
\label{fig:Ams2}
\end{figure*}

As discussed in Section \ref{sec:bin_res_early}, the Fourier $A_m$ values provide a global measure of the strengths of nonaxisymmetric density structures produced by GIs. Fig. \ref{fig:Ams2} shows the $\varA{time-averaged}$ Fourier amplitudes $\langle A_m \rangle$ over the time interval $t=$ 14$-$21 ORPs, calculated over the radial region 10$-$40 AU, as a function of $m$. While $A_m$ values give a measure of the strengths of the nonaxisymmetric structures, these should not be taken as necessarily representing the strengths of individual, independent spiral waves with $m$ arms. 
For nonlinear one, two, and $\varA{three-armed}$ waves, power is expected at all values of $m$. Therefore, an individual $m$ should not necessarily be considered as representing modes \citep[see discussion in][]{Michael12}. 

Figs \ref{fig:Ams} and  \ref{fig:Ams2} reveal noticeable differences in the amplitudes of structures with different $m$-symmetry, even in the $\varA{quasi-steady}$ state. The amplitude of the $\varA{one-armed}$ spiral of the circumbinary disc is higher than that of the control disc. On the other hand, after $t = 7$ ORPs, the GIs in the control disc are dominated by $m = 2$ over most of the simulation. As shown in \citet{MichaelT11}, the $m = 2$ Fourier component in the control disc is produced by a global two$-$armed spiral mode with corotation near 25 AU. The stronger $m = 1$ component in the circumbinary case produces somewhat stronger $m = 3$ and slightly lower $m = 2$ and $4$ in the asymptotic $\varA{quasi-steady}$ phase, as can be seen in Figs \ref{fig:midplane_2017}, \ref{fig:Ams} and in the inset of Fig. \ref{fig:Ams2}. Nevertheless, these structural differences do not much affect the $\varA{late-time}$ $Q$ distributions shown in Fig. \ref{fig:Qtoomre}.

\begin{figure*}
        \centering
                \includegraphics[width=\textwidth]{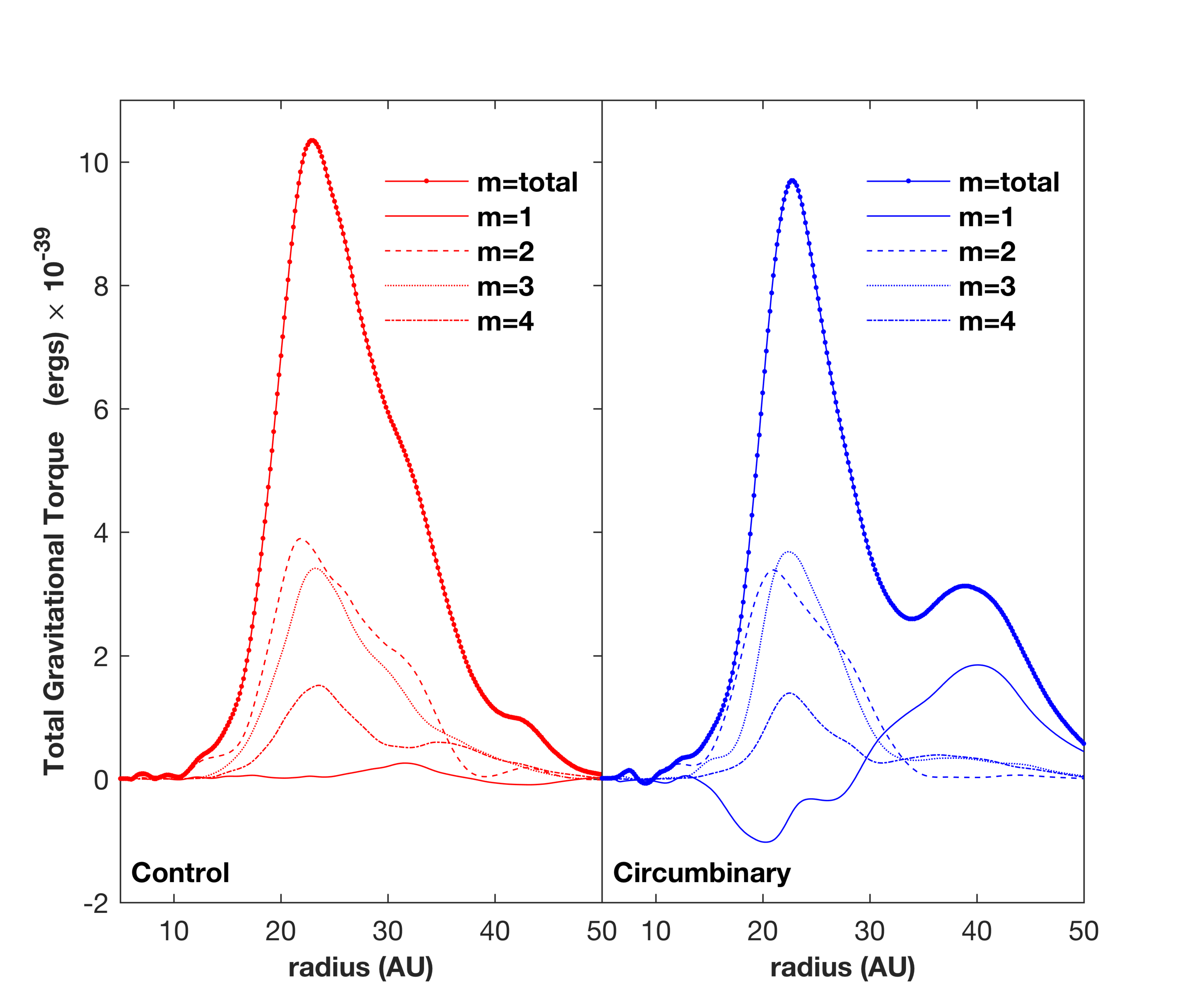}
  \caption{Gravitational torques $\varA{time-averaged}$ for 14 to 21 ORPs for individual $\varA{low-order}$ Fourier terms, and the total gravitational torque summed over all the Fourier terms. The plotted values do not represent local torques but rather the gravitational stresses integrated over a cylinder at radius $r$ (see text). Therefore, the slope of the torque curve determines whether local ring annuli are losing or gaining angular momentum.}
\label{fig:bin_torque}
\end{figure*} 
%
\subsubsection{Gravitational Torques}  
\label{sec:grav_tor}
%

The angular momentum and mass in this $\varA{GI-active}$ disc are transported mainly through gravitational stresses, as shown in \citet{Boley2006-2}, \citet{Michael12}, and \citet{SteimanCameron2013}. These papers explain in detail how we calculate the internal gravitational torques in the disc for different $m$-values by a Fourier decomposition, and they demonstrate that essentially all the mass transport in the $\varA{quasi-steady}$ phase is accounted for by these torques alone. In simulations by other researchers, {\em local} shearing disc simulations tend to yield comparable gravitational and Reynolds stresses \citep[e.g.,][]{Gammie2001,JohnsonGammie2003}, while measured contributions from Reynolds stresses are small in other {\em global} simulations \citep[e.g.,][]{LodatoRice2004}.

Fig. \ref{fig:bin_torque} presents the $\varA{time-averaged}$ gravitational torques for several $\varA{low-order}$ $m$-values along with the total gravitational torques arising from the sum of all the Fourier terms. The quantity plotted represents the torque exerted by the entire disc interior to $r$ on the entire disc exterior to $r$.  It is not a local torque but rather the gravitational stress integrated over a cylinder at $r$. The net torque on an annulus of the disc is the difference between the plotted values at the inner edge of the annulus and the outer edge of the annulus.  Therefore, it is the slope of the torque curve that determines whether local ring annuli are losing (positive slope) or gaining (negative slope) angular momentum.

Torques shown in Fig. \ref{fig:bin_torque} are $\varA{time-averaged}$ over 14 to 21 ORPs (a time interval of 1,300 years) because the torques fluctuate by about ten percent on orbital time scales. This should not be surprising given the fluctuations evident in $A_m$ values shown in Fig. \ref{fig:Ams}. Torques due to $\varA{two-armed}$ spirals are slightly larger in the control disc while torques due to $\varA{three-armed}$ spirals are slightly larger in the circumbinary disc. The circumbinary disc has larger torques for $m =1$ structures. The presence of the binary is clearly responsible for the increase. When the torques are summed over all Fourier components excepting $m = 1$, it is found that total torques in both discs are very similar.

The primary result in Fig. \ref{fig:bin_torque} is that, averaged over the entire $\varA{quasi-steady}$ phase, the torques due to structures with $m = 2$, $3$ and $4$ are remarkably similar for the two simulations, with small shifts in relative strengths of $m = 2$ and $m = 3$ corresponding to slight differences in $\langle A_m \rangle$ values in Fig. \ref{fig:Ams2}. The significant effect of the brown dwarf's presence at this late phase is a strong $m = 1$ torque that permeates the disc and actually introduces negative torques over $\sim$ $15-30$ AU that slightly reduce the peak torque there. At $r \gtrsim 30$ AU, the $m = 1$ torque component has a broad positive peak that produces a secondary peak in the total gravitational torque. Over time, this could produce substructure in the radial mass distribution, like an annulus of enhanced surface density in the region around 35 AU. This should occur because a positive slope in Fig. \ref{fig:bin_torque} tends to correspond to inward transport of mass, while a negative slope tends to correspond with outward transport of mass. As a result, with the $\varA{double-peak}$ profile of the binary case, it is possible for mass to accumulate somewhere interior to the outer peak.

The $m = 1$ torque is the most obvious qualitative difference caused by the binary companion in the quasi-steady phase and is fairly strong even after almost 4,000 years. Given that the $m = 1$ power declines slowly but steadily over the circumbinary simulation (Fig. \ref{fig:Ams}), it is unclear how long the $m = 1$  torque will influence a circumbinary disc. It could be just a decaying transient due to the initial conditions or it may be a permanent feature, perhaps a forced SLING-like phenomenon \citep{Shu1990} driven by the companion's presence. Longer simulations could shed more light on this.

%
\subsubsection{Radial Mass Transport}  
\label{sec:bin_res_mass}
%
\begin{figure*}
        \centering
               \includegraphics[width=\textwidth]{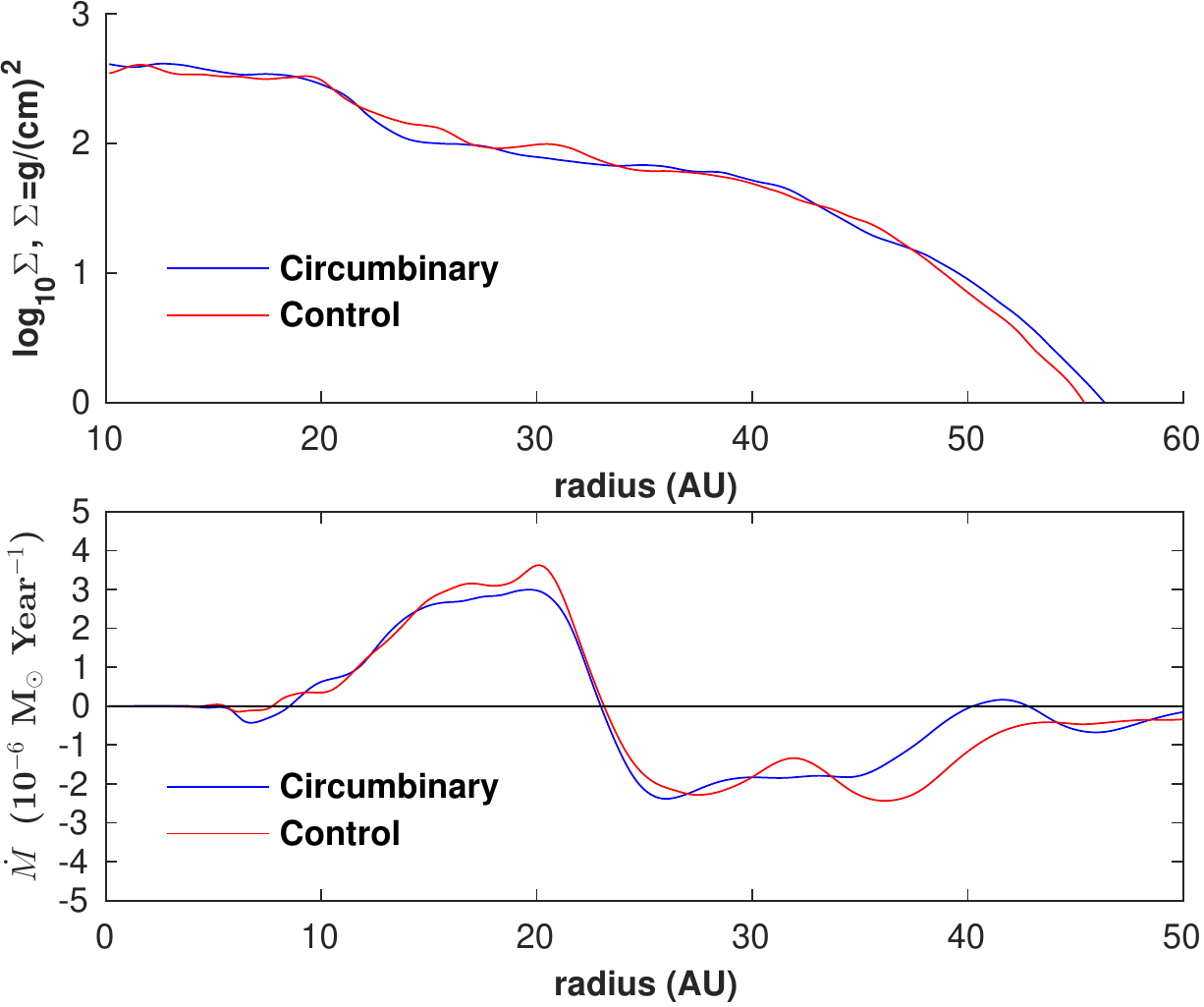}
\caption{Top: Azimuthally averaged surface density profiles for both discs at 21 ORPs, nearly at the end of the simulation. The overall density profiles are very similar for both discs. Bottom: Radial mass transport averaged over 14 to 21 ORPs in both discs.}
\label{fig:mdot}
\end{figure*}
As shown in the top panel of Fig. \ref{fig:mdot}, both discs settle into very similar azimuthally averaged surface density distributions. The initial $\varA{power-law}$ surface density distribution with $p = 0.5$ is relaxed to $p \sim1.5$ for the $10-40$ AU region. Even though there is continued mass transport in the $\varA{quasi-steady}$ phase, most of the density restructuring occurs during the violent $\varA{burst-like}$ onset of the GIs. It is remarkable that, despite the very different nature of the burst in these two simulations, the final surface density profiles end up quite similar.

The radial mass transport rate during the $\varA{quasi-steady}$ phase is shown in the bottom panel of Fig. \ref{fig:mdot}. Average mass fluxes are obtained by differencing the total mass inside cylindrical shells at two different times and dividing the result by the time interval. The rates are plotted out to a radial distance of 50 AU, because $\sim$99$\%$ of the disc mass is within the central 50 AU for both discs. 

Peak inflow occurs at $\sim20$ AU for both discs, with mass inflow rates $\approx$ 3 $\times$ 10$^{-6}$  M$_{\odot}$ per year. The mass flux transitions from inflow to outflow around 23 AU in both discs, with inflow interior to this radius and outflow exterior to this radius. The net result is disc spreading. Spreading is a continuous process as the discs evolve. Very active mass transport is observed in the region between 10 and 40 AU, and the gravitational torque values for this region are also higher than elsewhere. About $80 \%$ of the gravitational torque contribution in this region is from two, three and four$-$armed spirals in both discs (see Fig. \ref{fig:bin_torque}), demonstrating that these structures, as can also be seen in Fig. \ref{fig:midplane_2017}, are largely responsible for the mass transport. 

Overall, the mass flux profiles for the two simulations in the $\varA{quasi-steady}$ phase are quite similar. The one noticeable difference is at radii around 40 AU. A comparison of Fig. \ref{fig:bin_torque} and Fig. \ref{fig:mdot} shows that the difference in mass flow is associated with the secondary peak in the total gravitational torque in the circumbinary simulation arising from the $m = 1$ torque component. If this difference persists for $\sim$ 10$^{4}$ to 10$^5$ years or more, the circumbinary disc would develop an annular region of enhanced surface density around 40 AU that will not be present in the control disc.
This accumulation of mass would be due to the secondary torque peak, as discussed in Section \ref{sec:grav_tor}.
%
\subsection{Binary Orbit Evolution}  
\label{sec:bin_orbit}
%
 
\begin{figure*}
        \centering
                \includegraphics[width=\textwidth]{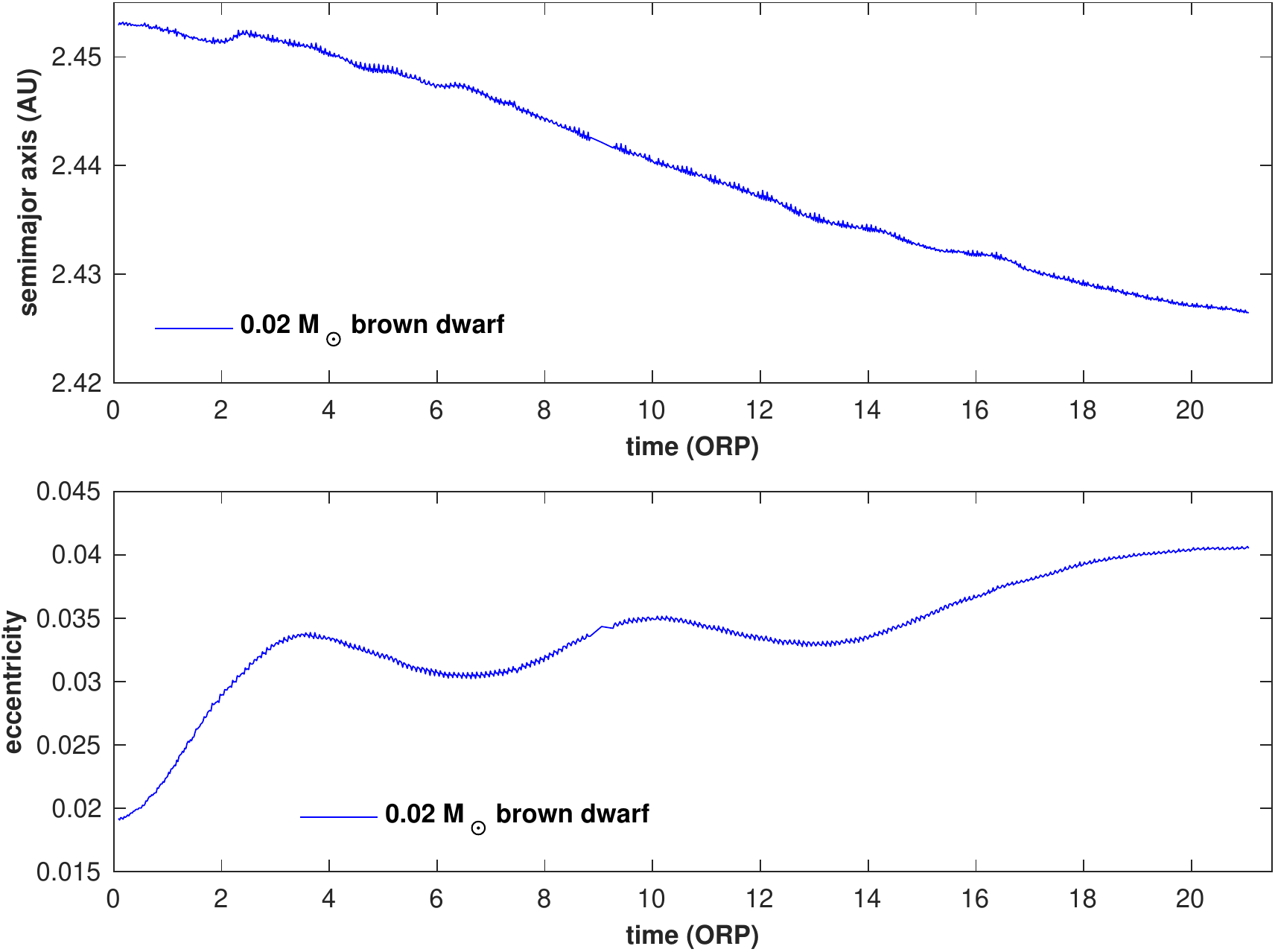}
\caption{Top: Semimajor axis of the binary's relative orbit as a function of time. Bottom: Eccentricity of the relative orbit plotted vs. time in ORPs. The gap near 9 ORPs is due to loss of a dataset.}
\label{fig:orbit}
\end{figure*}

Fig. \ref{fig:orbit} shows the $\varA{time-evolution}$ of the orbital eccentricity and the semimajor axis of the binary relative orbit. The average period for all the orbits of the binary is 0.021 ORPs. The local Keplerian period at the initial semimajor axis is 0.022 ORP. The eccentricity of the binary roughly doubles over the course of the simulation from the initial value of 0.019 to the final value of 0.040. Its average rate of increase is $\sim$ 0.001 per ORP $\approx$ 6 x 10$^{-6}$ per year while oscillating around that trend. If this growth were sustained for over 10$^5$ years, the eccentricity of the system could approach unity. The initial semimajor axis of the binary is 2.453 AU and the final semimajor axis is 2.426 AU. Over the 3,800 year duration of the simulation, the semimajor axis thus decreases by about one percent. Again this suggests significant orbital evolution through interaction with the disc if the simulation were extended to 10$^5$ years.

\subsection{Inner Disc Vortex}  
\label{sec:vortex}
%

\begin{figure*}
        \centering
                \includegraphics[width=\textwidth]{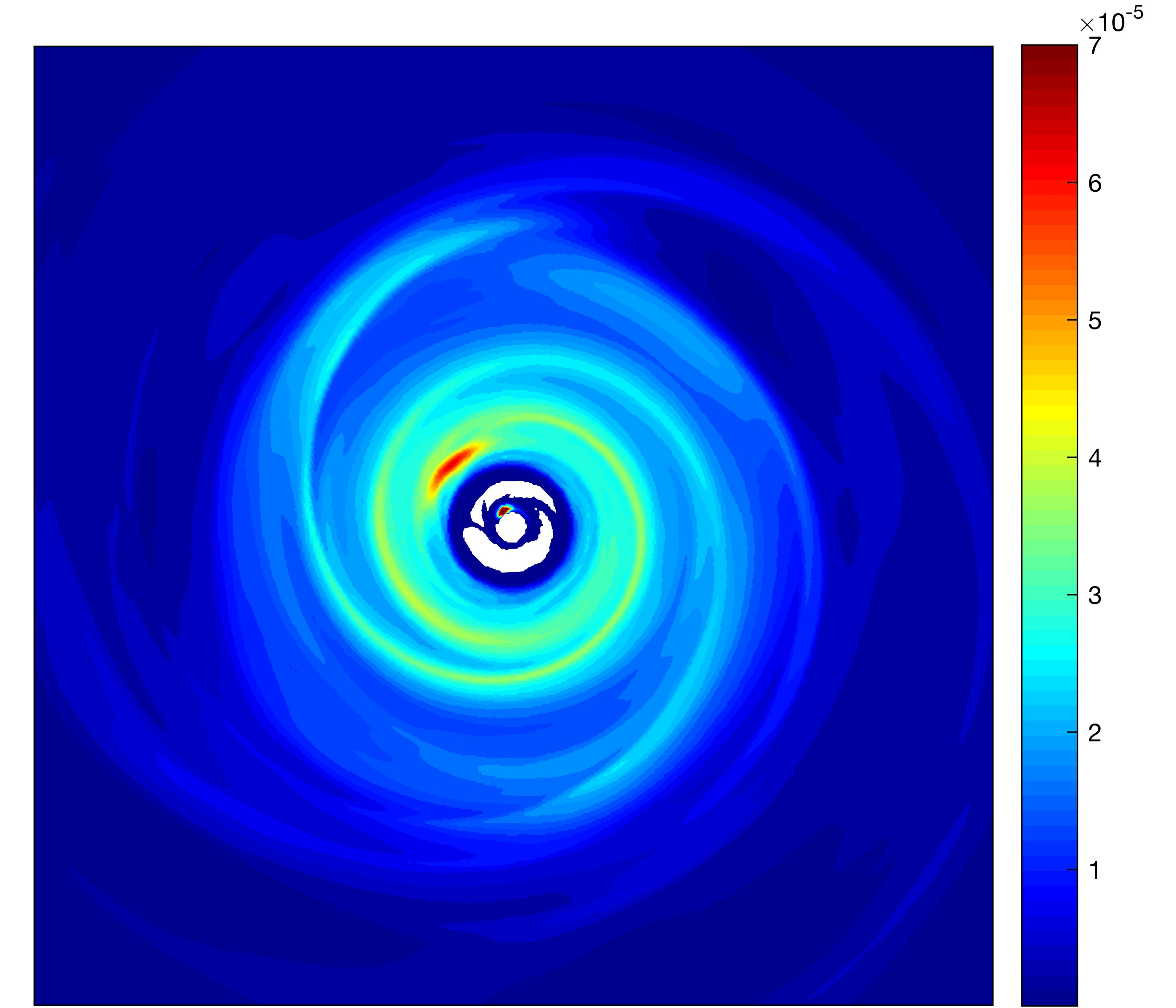}
\caption{100 AU by 100 AU plot of the midplane density of the circumbinary disc at 21 ORPs on a logarithmic scale. Gas orbital motion in the disc is counterclockwise. Anticyclonic (clockwise) flows are found around the red vortex in the inner disc when viewed in the frame of vortex center.}
\label{fig:vort}
\end{figure*}

Our analysis of midplane density structures reveals a persistent $\varA{over-density}$ in the inner regions of both discs. These $\varA{banana-shaped}$ features appear to be vortices.  In fact, the velocities near the density maxima show anticyclonic motion, as expected for a $\varA{long-lived}$ vortex in a nearly Keplerian disc \citep{Adams1995}. The primary focus of this work is to understand how the central binary affects the evolution of the gravitational instabilities in the bulk of the protoplanetary disc away from the inner edge.  However, we want to acknowledge an indication of vortex formation in $\varA{GI-active}$ discs because it could be of interest to the community.  Because of limitations in the vertical resolution of the inner disc in our simulations, we cannot rule out the possibility that these vortices are numerical artifacts \citep{Boley2007a}, but they do behave the way a vortex is expected to behave.  
 
The red oval$-$shaped feature in Fig. \ref{fig:vort} is the vortex in the circumbinary disc at 21 ORPs. The density structure of the vortex in the control disc is very similar. Both vortices have centers located at radii of $\sim$ 6 to 7 AU and are $\varA{long-lived}$. They appear at 4 ORPs and persist through the end of the simulations at a roughly constant density contrast with their surroundings. Their longevity and their constant density structure indicate that the vortices are stable features in the inner disc. 

The vortices do vary slightly in strength throughout the simulations. The vortex observed in the control disc is not as robust as the one observed in the circumbinary disc and seems to be dissipating by 21 ORPs. The orbital periods of the vortex centers are 0.09 ORP, essentially the same as the disc rotation period at these radii. The mass of the vortex in Fig. \ref{fig:vort} is estimated to be $1 \pm 0.5 $ M$_{\jupiter}$. The precise mass of the vortex is difficult to determine due to uncertainties inherent in defining the vortex boundaries.

If physical and not numerical in origin, the vortices in both discs are probably due to a Rossby wave instability (RWI) caused by inner edges of discs \citep{Lovelace99,Li00,Li01}. RWI has been shown to be responsible for the formation of vortices in protoplanetary discs \citep[][and references therein]{Lovelace_2014}. In this paper, we only present some relevant details of the vortex.  Proper treatment of the vortex and a detailed analysis of the disc's susceptibility to the RWI would require a simulation with significantly increased resolution of the inner disc, which goes beyond the scope of this paper.

%
\section{Discussion}  
\label{sec:discussion}
%
The reported simulations are most relevant to the early evolution of protoplanetary discs because the 0.14 M$_{\odot}$ disc is relatively massive. This work provides a comparison between GIs in a marginally unstable radiatively cooling disc around a single star and the same disc orbiting a close binary of low mass ratio contained inside the central hole of the disc. 

%
\subsection{GIs in both Discs}  
\label{sec:bin_disc_GI}
%
The initial conditions in both simulations are artificial. There are two reasons why the onset of GIs are different during the first 10 ORPs: (1) $\varA{one-armed}$ disturbances are much stronger in the circumbinary disc and (2) the initial disc is significantly out of equilibrium once the brown dwarf is included. These two effects are probably not independent. The $\varA{one-armed}$ spiral is probably induced at least in part by the initial deviation from equilibrium. In hindsight, the mass of the central star in the circumbinary disc simulation should have been reduced to 0.98 M$_{\odot}$, which would have conserved the total mass of the binary system as 1 M$_{\odot}$, equal to the mass of the primary in the control disc. Having an equal amount of mass in the inner hole in both models would have provided a somewhat better comparison. Still, the initial disc would be out of equilibrium when the central object is split into two masses even if the total mass is conserved.
 
Both discs are in a similar stage of sustained marginal instability in the $\varA{quasi-steady}$ phase by 14 ORPs. Strengths of the global nonaxisymmetric density features are also similar in both discs. Subtle differences exist in the nonaxisymmetric power and torques due to $m = 2$, 3 and 4 armed structures. There is, however, a strong $m = 1$ difference lingering even in the $\varA{quasi-steady}$ phase. Torques due to $m =1$ are much stronger and have an entirely different radial profile in the circumbinary disc than in the control disc. Our simulations show that a disc like this, with radiative cooling times typically much greater than the local dynamic time throughout the disc \citep{Boley2006-2,SteimanCameron2013}, does not fragment but seeks out a marginally unstable state of balance between GI heating and radiative cooling regardless of a significant difference in GI onset and perturbations by a central binary of low mass ratio.

The presence of the binary companion leads to excess heating in the inner disc. We do not have enough resolution in the inner disc to be certain, but it is likely that this is due to shocks produced in the inner disc through gravitational perturbations by the brown dwarf. Heating in binary protoplanetary discs can prevent formation of planetesimals \citep{Nelson2000,Mayer2005}. Similarly, higher temperatures in the inner region of the circumbinary disc can inhibit the planetesimal formation process.

%
\subsection{Comparison with Other Research}  
%

There are two competing effects when the tidal forces due to binarity are strong. This has been studied mostly for discs around the individual stars in a binary. Even in this case, results remain somewhat uncertain. On the one hand, strong spiral arms induced by tidal forces produce strong shocks that heat the disc, which has the effect of raising $Q$ and suppressing disc fragmentation \citep{Nelson2000}. On the other hand, strong perturbations can lead to fragmentation if a disc cools quickly enough  \citep{Boss06bin,Mayer_bin_2010}. 
 
\citet{NelsonMarzari2016} simulated circumbinary tori and discs meant to resemble the GG Tau system. Their simulations did produce fragmentation, but there was no comparison $\varA{single-star}$ simulation to judge the effects of binarity. 

Our simulations using the rigorously tested radiative cooling scheme of \citet{Boley2006-2,Boley2007a} demonstrate that a binary of low mass ratio (0.02) placed inside a disc with a long cooling time does not make a large overall difference to the asymptotic $\varA{quasi-steady}$ behavior of the GIs. We suspect but cannot prove that this would remain true even if the mass ratio were increased to $\sim$ 0.1. On the other hand, the onset of GIs, if initiated by a burst, can be significantly different with and without a binary. We would expect this difference to be more pronounced as the binary mass ratio were increased. A $\varA{one-armed}$ spiral produced by a GI burst in a circumbinary disc should be readily visible in both continuum and molecular line emission \citep{Ilee2011,Douglas2013,Evans2015,Evans2017}.

\citet{Mutter2017,Mutter2017_2} performed 2D hydrodynamics simulations of the $\varA{Kepler-16}$, -34, and -35 systems to investigate the role of the disc's $\varA{self-gravity}$. They find that GIs can have a significant impact on the formation and evolution of circumbinary planets in these systems. In their simulations, a pronounced $m = 1$ global spiral wave is seen due to the disc $\varA{self-gravity}$. Their simulated orbits agree with the $\varA{Kepler-16b}$ system, but the orbits of $\varA{Kepler-34}$ and -35 systems are difficult to explain.  
%
%
\subsection{Mass Transport}  
\label{sec:bin_disc_mass}
%
Gravitational torques in both discs are similar in the $\varA{quasi-steady}$ phase, with peak mass inflow rates of about 3 $\times$ 10$^{-6}$  M$_{\odot}$ per year. In both cases, when the GIs become nonlinear during the burst phase there is a significant mass redistribution within a few ORPs. The only substantial difference at late times is the additional $m = 1$ torque in the outer circumbinary disc, which 
alters the increased mass flow. 
If the excess $m = 1$ torques persist long enough, $\sim$ 10$^{4}$ to 10$^5$ years, they may form an enhanced density ring in the outer disc, which could be a site for planet formation in the outer disc through the accumulation of $\varA{meter-sized}$ particles at the radial pressure maximum \citep{Haghighipour2003}.

%
\subsection{The Inner Vortex and Planet Formation}  
\label{sec:bin_disc_vortex}
%

Higher resolution is required for a better understanding of the inner disc vortices. Because the vortex in the circumbinary disc persists from $\sim$4 ORPs to the end of the run, it could be a favorable site for solids to accumulate and possibly even form planetesimals \citep{Lyra2009_01}. \citet{Barge_1995} suggested that anticyclonic vortices in an accretion disc can aid the concentration of dust particles, which could facilitate the formation of planetesimals. The presence of a binary companion, while not responsible for the origin of the vortex, seems to play some role in maintaining its strength. 

Recent ALMA observations have suggested that a vortex in the inner disc can be an ideal site for accumulation of $\varA{millimeter-sized}$ dust particles \citep{Pinilla2015,Raettig15}, which can ultimately grow into planetary cores. Multifrequency observations of HD 142527 have shown crescent shaped regions capable of trapping $\varA{submillimeter-sized}$ particles in the pressure maxima \citep{Casassus2015}. Thus, a $\varA{GI-active}$ disc with vortices can be a candidate site for accumulation of dust that can form planetesimals. If the inner vortex is sustained long enough, this could lead to the formation of planetary cores.
 
%
\subsection{Future Work}  
\label{sec:future}
%

Onset of GIs is hastened by the presence of the companion, but no fragmentation due to GIs is observed in either simulation. Because the binary mass ratio in this work is only $0.02$ and the disc is 0.14 M$_{\odot}$, the effects of the companion on the GIs in the disc are not pronounced. More simulations involving a diverse set of the stellar and companion masses are needed to better understand GIs in circumbinary discs.  With the advent of ALMA, and with its ability to better resolve spiral structures in Class 0$-$1 discs, the role of GIs in the early evolution of protoplanetary discs will hopefully be better understood. Current observations of such discs are sparse \citep{Kratter16}. 
 
%
\section{Conclusion} 
\label{sec:conclusion}
%
The circumbinary simulation addresses the effects of a binary companion on the evolution of a young protoplanetary disc. A $\varA{low-mass}$ companion, a brown dwarf in this case, is not able to significantly alter the $\varA{long-term}$ behavior of GIs in a protoplanetary disc once they have reached a $\varA{quasi-steady}$ state; but it affects density structures and torques associated with the $m = 1$ symmetry. 
The increase in mass transport in some regions in the disc due to the $\varA{one-armed}$ wave could lead to the formation of an enhanced density ring, which could be a site of planet formation. The companion seems to stabilize the inner disc vortex. This could enhance the process of $\varA{particle-trapping}$ and ultimately planet formation. The orbital evolution of the binary is measured. In about 3,800 years, its semimajor axis decreases from the initial 2.453 AU to the final 2.426 AU and its eccentricity increases from the initial 0.019 to the final 0.040. If these trends persist over $\sim$ 10$^{5}$ years, the relative orbit would become significantly smaller and highly eccentric.

Acknowledgments:
We thank the anonymous referee for providing useful suggestions that improved the paper. This research was supported in part by Lilly Endowment, Inc., through its support for the Indiana University Pervasive Technology Institute and for the Indiana MAETACyt Initiative. This work was also supported in part by the NASA Origins of Solar Systems grant NNX08AK36G. This material is based upon work supported by the National Science Foundation under Grant No. CNS-0521433 and CNS-0723054. 

\bibliography{master}


\end{document}